\documentclass[9pt,twocolumn,twoside]{osajnl}

\journal{josab} 

\setboolean{shortarticle}{false} 
\usepackage{graphicx}
\usepackage[space]{grffile}
\ifthenelse{\boolean{shortarticle}}{\colorlet{color2}{color2b}}{\colorlet{color2}{color2}} 
\title{Comparison of collimated blue light generation in ${}^{85}$Rb atoms via the D${}_1$ and D${}_2$ lines}

\author[1,*]{Nikunj Prajapati}
\author[2,3]{Alexander M. Akulshin}
\author[1]{Irina Novikova}

\affil[1]{Department of Physics, College of William $\&$ Mary, Williamsburg, Virginia 23187, USA}
\affil[2]{Centre for Quantum and Optical Science, Swinburne University of Technology, Melbourne, Australia}
\affil[3]{Johannes Gutenburg University, Helmholtz Institute, D-55128, Mainz, Germany}
\affil[*]{Corresponding author: nprajapati@email.wm.edu}

\dates{Compiled \today}

\ociscodes{(020.4180) Multiphoton processes; (270.1670) Coherent optical effects.}

\doi{\url{http://dx.doi.org/10.1364/ao.XX.XXXXXX}}

\begin{abstract}
We experimentally studied the characteristics of the collimated blue light (CBL) produced in ${}^{85}$Rb vapor by two resonant laser fields exciting atoms into the $5D_{3/2}$  state, using either the $5P_{1/2}$ or the $5P_{3/2}$ intermediate state. We compared the CBL output at different values of frequency detunings, powers, and polarizations of the pump lasers in these two cases, and confirmed the observed trends using a simple theoretical model. We also demonstrated that the addition of the repump laser, preventing the accumulation of atomic population in the uncoupled hyperfine ground state, resulted in nearly an order of magnitude increase in CBL power output. Overall, we found that the $5S_{1/2} - 5P_{1/2} - 5D_{3/2}$ excitation pathway results in stronger CBL generation, as we detected up to $4.25~\mu$W using two pumps of the same linear polarization. The optimum CBL output for the  $5S_{1/2} - 5P_{3/2} - 5D_{3/2}$ excitation pathway required the two pump lasers to have the same circular polarization, but resulted only in a maximum CBL power of $450$~nW.
\end{abstract}

\setboolean{displaycopyright}{true}

\begin{document}

\maketitle
\thispagestyle{fancy}

\ifthenelse{\boolean{shortarticle}}{\ifthenelse{\boolean{singlecolumn}}{\abscontentformatted}{\abscontent}}{}

\section{Introduction}

Multi-photon processes have become an important tool in nonlinear and quantum optics for non-classical light and entanglement generation in a wide spectral range. While traditionally nonlinear crystals are used for frequency conversion and wave mixing, a strong nonlinearity can be also achieved in the proximity of optical transitions in atoms, reducing required laser power and removing the requirement of an optical cavity~\cite{bachor_guide_2004}. A broad variety of interaction configurations have been considered and realized for numerous applications. Among them, the scheme involving two-photon excitation reaching higher energy levels has been investigated for efficient frequency up-conversion~\cite{boydPRA87blue,garrettPRL93,lvovskiPRL99}, single-photon frequency conversion~\cite{radnaevNPhot10}, quantum memory~\cite{NunnORCApreprint}, selective non-linearity suppression~\cite{PrajapatiJOSAB17}, quantum noise dynamics~\cite{sautenkovPLA12blue}, etc. 

An excitation to the $nD$ states of alkali metal atoms opens interesting possibilities for nonlinear optics, 
as the population inversion, guaranteed between certain excited levels with appropriate lifetimes and branching ratios, results in amplified spontaneous emission (ASE) and spontaneously-seeded four-wave mixing for the involved optical transitions. A lot of attention was recently given to the generation of the collimated blue light (CBL) at $420.3$~nm via the $5S_{1/2} \rightarrow 5P_{3/2} \rightarrow 5D_{5/2}$ transition in Rb vapor~\cite{zibrov02pra,AkulshinOE09,VernierOE10,BrekkeOL13,bluelightAJP2013,SellOL14,AkulshinOL14,BrekkeOL15}. Such interacting systems have been successfully used to study the interplay of co-existing nonlinear processes~\cite{MeijerOL06,LeeOE16}, the effects of externally-seeded optical fields~\cite{AkulshinJOSAB17}, and of optical resonators~\cite{BrekkeAO17}. It also served as a tool for studies on orbital angular momentum conservation and manipulations in nonlinear processes~\cite{WalkerPRL12,AkulshinOL15,AkulshinOL16}.

We report on the investigation of CBL generation in the two-photon transition reaching the $5D_{3/2}$ state, providing an opportunity to investigate the interference between competing excitation channels, spontaneous decays, and nonlinear processes. It allows for an alternative, more symmetric, four-wave mixing diamond scheme involving only near-IR optical 
fields~\cite{PhysRevLett.111.123602,PhysRevLett.113.163601,PhysRevA.91.023418,LeszczynskiOE17}. Since it is important to understand the interplay between the multiple excitation and relaxation channels to maximize the efficiency of a particular nonlinear process, here we experimentally compared the two excitation pathways to the $5D_{3/2}$ level through either $5P_{1/2}$ or $5P_{3/2}$ intermediate levels, and identified the optimal conditions for the collimated blue light generation in each case.


\section{Experimental arrangements}

\begin{figure}[H]
	\centering
	\includegraphics[width=1\columnwidth]{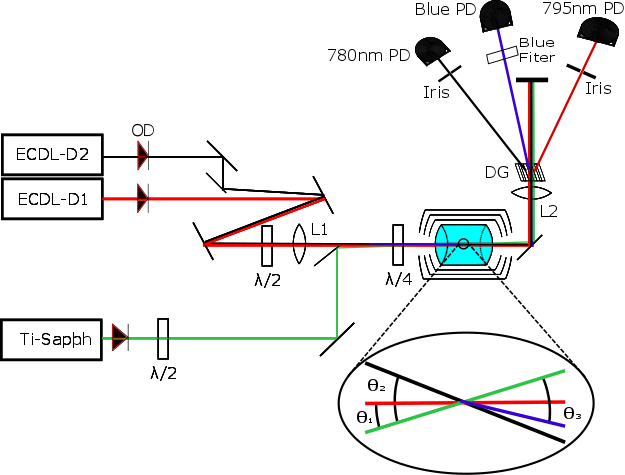}
	\caption{The optical layout of the experimental setup. ECDL-D1, ECDL-D2, and Ti:Sapph denote three independent lasers used in the experiment. The optical paths of the D1, D2, and Ti:sapph pump lasers and the generated blue light are show in, red, black, green, and blue, respectively. Inset shows relative orientation of the optical beams. 
	See text for the abbreviations.}
	\label{fig:setup}
\end{figure}

The schematic of the experimental apparatus is shown in Fig.~\ref{fig:setup}. We employed three individual lasers. Two external cavity diode lasers -- ECDL-D1 and ECDL-D2 -- tunable in the vicinity of the Rb $D_1$ line (wavelength $795$~nm) and Rb $D_2$ line (wavelength $780$~nm). Each ECDL, depending on the stage of the experiment, can serve as either lower pump or re-pump laser while the upper pump optical field is generated using the continuous wave (cw) Titanium Sapphire (Ti:Sapph) laser. The first stage utilizes the D1 laser as the lower pump, the D2 as the re-pump, and the Ti:Sapph tuned to 762~nm (for $5P_{1/2}\rightarrow5D_{3/2}$ transition) while the second stage involves the D1 and D2 lasers swapping roles and the Ti:Sapph being tuned to 776~nm (for the $5P_{3/2}\rightarrow5D_{3/2}$ transition).

The two fields generated by the ECDLs were combined first so they could be adjusted together before combining with the Ti:Sapph laser. In order to further increase the laser intensities, the laser beams were weakly focused inside the Rb cell using a 1000~mm (L1) lens and then collimated using a 500~mm (L2) lens. All beams had gaussian intensity profiles with diameters $230~\mu$m, $250~\mu$m, and $840~\mu$m at the center of the Rb cell, for the D1, D2, and Ti:Sapph laser beams, respectively.

For maximum flexibility in setting the pump field polarizations, all optical fields were combined using edge mirrors. We found that the polarization of the re-pump field relative to the lower pump field had very little effect on CBL generation, and thus we always matched the repump laser polarization to that of the lower pump field. The polarizations of the lower and upper pump fields, before entering the cell, were controlled independently using half- and quarter -wave plates. The polarizations of the indavidual beams were cleaned using beam splitters before they were combined. However, polarization imperfections could have risen from the use of zero-order waveplates designed for $795$~nm light.


In the experiment we used a 75~mm - long cylindrical Pirex cell (diameter 22~mm), containing isotopically enriched $^{85}$Rb vapor. The cell was tilted by approximately $6^\circ$ to avoid the retroreflection effects from the cell's windows on the generated CBL~\cite{AkulshinJOSAB17}. For all the measurements the cell was maintained at a relatively low temperature of $88^oC$, corresponding to the $^{85}$Rb density of  $ \approx 1.7x10^{12}~\mathrm{cm}^{-3}$. The cell was housed in three layer magnetic shielding, with the innermost layer wrapped in a heating wire. Thermal insulation was placed between each layer of the magnetic shield to help with temperature stability.

Under these conditions we observed the emergence of collimated blue light.  To maximize its power, we adjusted the relative angles between the two co-propagating pump laser fields and the repump laser as shown in the inset of the Fig.~\ref{fig:setup}: all three beams were arranged in the same plane, with the angles between the Ti:Sapph laser and D1 and D2 laser beams being $\theta_1=2.1$~mrad and $\theta_2=7.5$~mrad correspondingly. The output CBL beam then emerged at the angle of $\theta_3=3.3$~mrad from the Ti:Sapph beam. We found that for both intermediate $5P$ states, the generated blue light was produced at a wavelength of $421.7$~nm (measured using an Ocean Optics spectrometer with spectral resolution $\pm 0.2$~nm) corresponding to the $6P_{1/2} \rightarrow 5S_{1/2}$ optical transition. We were not able to detect any directional radiation at the $5D_{3/2} \rightarrow 6P_{1/2}$ and $5D_{3/2} \rightarrow 6P_{3/2}$ transitions, since the glass cell is not transparent in the mid-IR spectral range. We also did not observe optical fields corresponding to the alternative relaxation pathways through the $6S_{1/2}$ state ~\cite{AkulshinOL14,SellOL14,AkulshinOL15} or $5P$ states~\cite{PhysRevA.91.023418,LeszczynskiOE17}.

To separate the CBL beam from the pump fields after the Rb cell, the output beams passed through a diffraction grating (DG) which directed $\approx 46~\%$ of the total power of each field into the first diffraction order. We then used irises to isolate individual laser fields before the photodetectors (PD). To avoid contamination of the CBL measurements by any scattered IR laser light, we placed a blue spectral filter (transmission $\approx 40\%$ at 421.7~nm) before the corresponding photo-detector.

\section{CBL generation via the $D_1$ line}

\begin{figure}[H]
	\centering
	\includegraphics[width=.5\columnwidth]{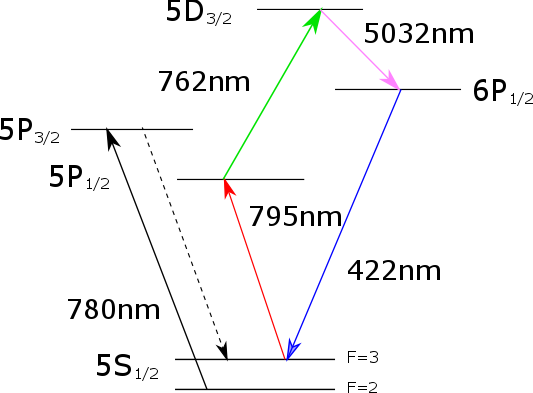}
	\caption{Interaction configuration through the $5P_{1/2}$ intermediate level: lower pump ($794.97$~nm) and the upper pump ($762.103$~nm) excite Rb atoms to the $5D_{3/2}$ level, followed by the emission of 5.032~$\mu$m (not detectable) and collimated blue light at 421.7~nm. The repump field is tuned to the $5S_{1/2}, F=2 \rightarrow 5P_{3/2}$ transition.
 }
	\label{fig:levels_D1}
\end{figure}

In this section we present the measurements in which the $D_1$ transition ($5S_{1/2} \rightarrow 5P_{1/2}$, $\lambda_{D1} = 795$~nm) served as the first step of the excitation scheme; the second pump laser with the wavelength 762~nm was used to further excite atoms into the $5D_{3/2}$ excited state, as shown in Fig.\ref{fig:levels_D1}. In this configuration the $D_2$ laser, acting as a repump, was tuned to the transition between the excited $5P_{3/2}$ level and the ground-state hyperfine sublevel, not coupled by the $D_1$ pump laser ($F=2$ in this case). Unless otherwise noted, all the reported data are recorded with the repump laser on, as it produced a uniform increase in the recorded CBL power, regardless of polarizations and powers of the pump lasers. 

\begin{figure}
	\centering
	\includegraphics[width=.8\columnwidth]{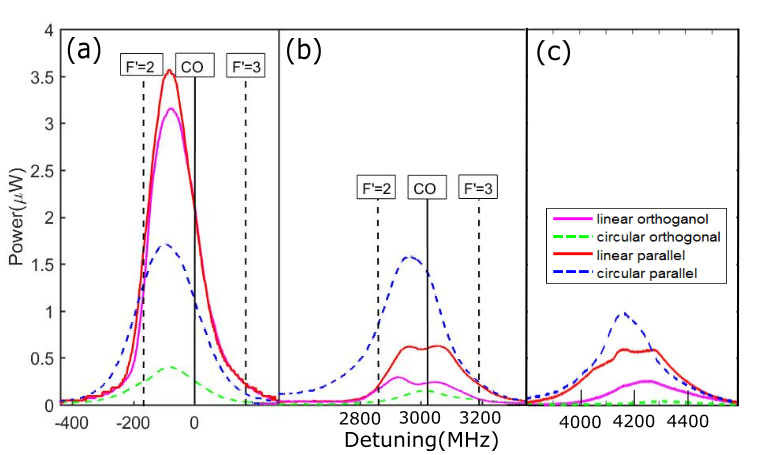}
	\caption{Power of the generated blue light as the lower pump was swept across: \emph{(a)} $5S_{1/2}, F=3\rightarrow 5P_{1/2} $ transition,  \emph{(b)} $5S_{1/2}, F=2\rightarrow5P_{1/2}$, and \emph{(c)} $1.2$~GHz above $5S_{1/2}, F=2\rightarrow5P_{1/2}$ transitions. The upper pump was tuned to 762.1036~nm for \emph{(a,b)}, and 762.1054~nm for \emph{(c)}. Four different polarization configurations of the two pump lasers are shown: linear parallel (red, solid line), linear orthogonal (magenta, solid line), circular parallel (blue, dashed line), and circular orthogonal (green, dashed line). The powers of both lower pump ($D_1$ laser) and the repump ($D_2$ laser) were kept at 16~mW,  and the power of the upper pump (Ti:Sapph laser) at 200~mW . The zero detuning of the $D_1$ pump corresponds to the cross-over transition of the $5S_{1/2}, F=3 \rightarrow 5P_{1/2}$ state.}
	\label{fig:polarization}
\end{figure}

CBL generation was analyzed for four pump polarization configurations, in which the two pump fields had either parallel or orthogonal linear or circular polarization. The resulting observations are shown in ~Fig.\ref{fig:polarization}, in which we plotted the CBL power for each polarization configuration as the function of the lower pump frequency (the upper pump frequency was fixed). We have considered three cases in which the lower pump laser was scanned across each hyperfine transition of the $D_1$ line [Fig.\ref{fig:polarization}(a,b)], as well as when it was detuned by $\approx +1.2$~GHz from the  $5S_{1/2}, F=2\rightarrow5P_{1/2} $ [Fig.\ref{fig:polarization}(c)]. For the resonance cases we found that the maximum blue light power corresponds to the frequency of the maximum $D_1$ laser absorption.

We found that the polarization configuration leading to the maximum blue light generation was different, depending on the laser frequency. We detected the strongest CBL generation at the lower frequency transition ($5S_{1/2}, F=2,3\rightarrow5P_{1/2} $)  when the two pump field were linearly polarized, with parallel arrangement results in slightly higher CBL power. However, the circularly polarized pump fields produced a similar amount of blue light for both the lower and higher-frequency transitions (5$S_{1/2}, F=3\rightarrow 5P_{1/2}$), but since the blue output for the linearly polarized pumps dropped significantly in the latter case, the circular parallel pumps maximized the CBL generation. Finally, the circular orthogonal configuration led to the smallest generation of CBL.

We also analyzed the CBL polarization for linearly polarized pump lasers. We found that for all investigated laser detunings the polarization of the generated blue light matched the polarization of the lower pump field, even for orthogonally polarized laser fields. Unfortunately, we were not able to carry out the CBL polarization analysis for the circularly polarized pumps since a quarter-wave plate for blue light unavailable.


\subsection*{On-resonant $D_1$-line excitation}

To investigate the power dependence of the generated blue light on all three involved laser fields, we considered on- and off-resonant tuning of the pump fields. In the first case, both pump fields were tuned near the centers of the corresponding optical resonant absorption peaks ($5S_{1/2}\rightarrow 5P_{1/2}$ and $5P_{1/2}\rightarrow 5D_{3/2}$). As CBL is the product of parametric wave mixing of two pump laser fields and the internally generated mid-IR field~\cite{AkulshinOE09,LeeOE16}, the maximum of the blue spectral profile did not always occur exactly at the two-photon resonance (in which the sum of the two laser frequencies exactly matched the frequency difference between the ground state and the excited $D$ state), but was shifted toward the frequency corresponding to the maximum lower pump absorption and often resembled two poorly-resolved peaks. For the power dependence studies we chose the lower pump detuning near the $5S_{1/2}, F=3\rightarrow 5P_{1/2}$ transition and parallel linearly polarized pump fields which produced the highest CBL output.                                                                                                                                                                                                                                                                                                                                                                                                                                                                                                                                                                                                                                                                                                                                                                                                                                                                                                                                                                                                                                                                                                                                                                                                                                                                                                                                                                                                                                                                                                                                                                                                                                                                                                                                                    

\begin{figure}[H]
	\centering
	\includegraphics[width=.7\columnwidth]{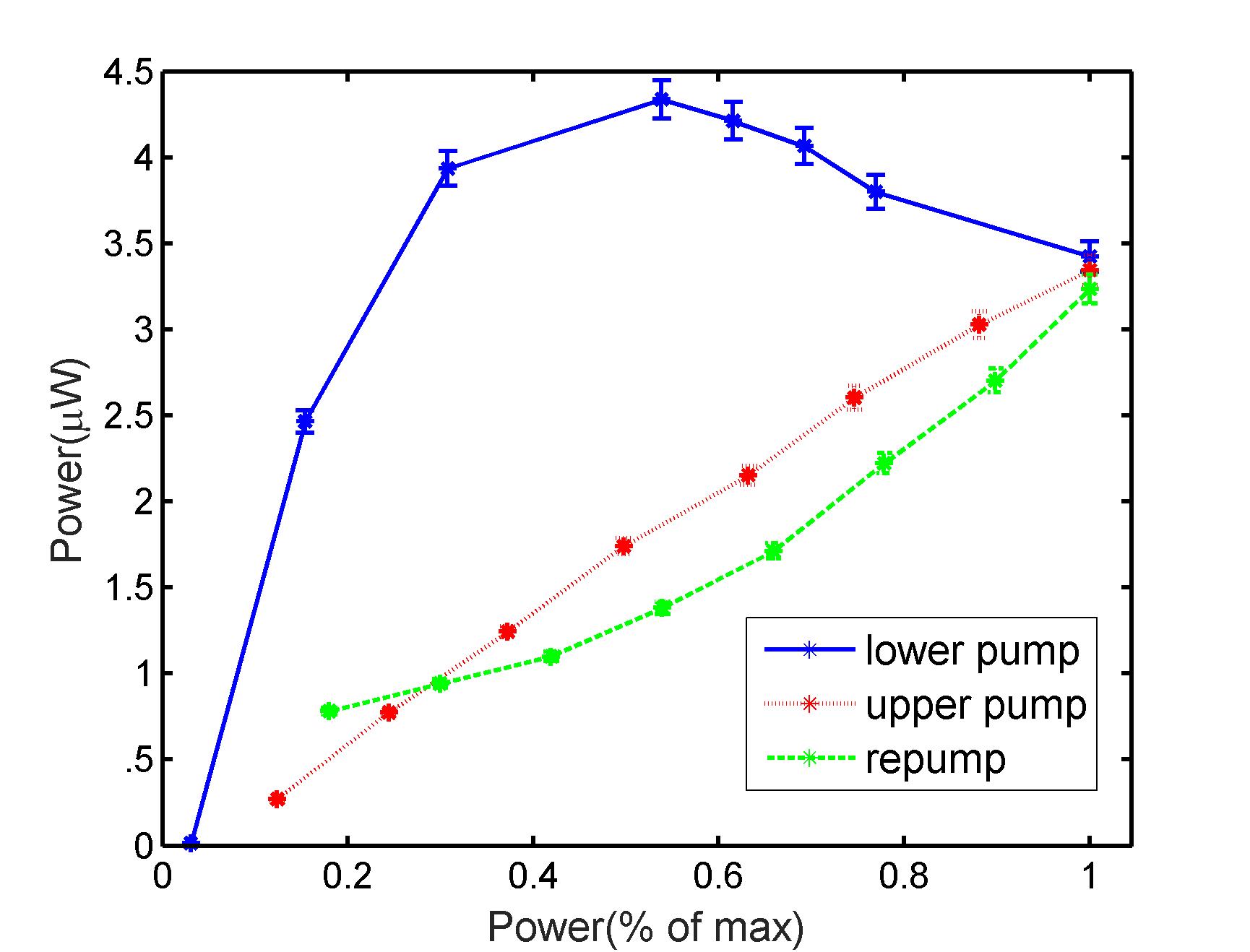}
	\caption{Generated CBL power as a function of normalized power of each pump and repump fields. For each individual dependence the power of one laser was varied between zero and its maximum value, while the other two lasers were kept at their maximum powers: 65~mW for the lower pump (795~nm), 200~mW for the upper pump (762~nm), and 17~mW for the repump laser (780~nm). The laser detuning corresponded to the conditions for the maximum CBL power as shown in Fig.~\ref{fig:polarization}(a).}
	\label{fig:D1_blue_on_res}
\end{figure}

At maximum power for all three fields, we measured $4.25~\mu$W of the generated blue light. As we decreased the power of the upper pump field, the CBL power dropped more or less linearly, as expected for the optically-driven population of the $5D_{3/2}$ excited state~\cite{GrovePhScripta95}. The reduction of the repump power resulted in a similar nearly linear drop in CBL until leveling off at 30$\%$ of the repump power. It is likely that the effect of the repumping became negligible for lower repump laser powers due to its strong absoprtion, since the  measured CBL power output ($500-700$~nW) matched the blue light generated in the complete absence of the repump field. 

However, the lower pump power dependence is more complicated: as the $D_1$ laser power increases, the CBL power grew steadily until it reached its plateau at $4.25~\mu$W at $\approx 50\%$ of the maximum available laser power ($\approx 30$~mW), and then began slowly decrease. The origin of such behavior is related to the optimization of excitation and relaxation rates from and to the ground state via stimulated processes, as will be discussed later in Sec.~\ref{sec:theory}. We have verified that the resonant absorption of the $D_1$ laser field displayed no similar trends, steadily decreasing from $70\%$ to $40\%$ with the growing laser power. 
It also should be noted that the reduction in the generated CBL power at higher pump power occurred only when the repump field was present. Without the repump, the CBL reached saturation at the $D_1$ field power of $\approx 35$~mW, and then stayed roughly at the same level with further laser power increase.



\subsection*{Off-resonant $D_1$-line excitation}

\begin{figure}[H]
	\centering
	\includegraphics[width=.7\columnwidth]{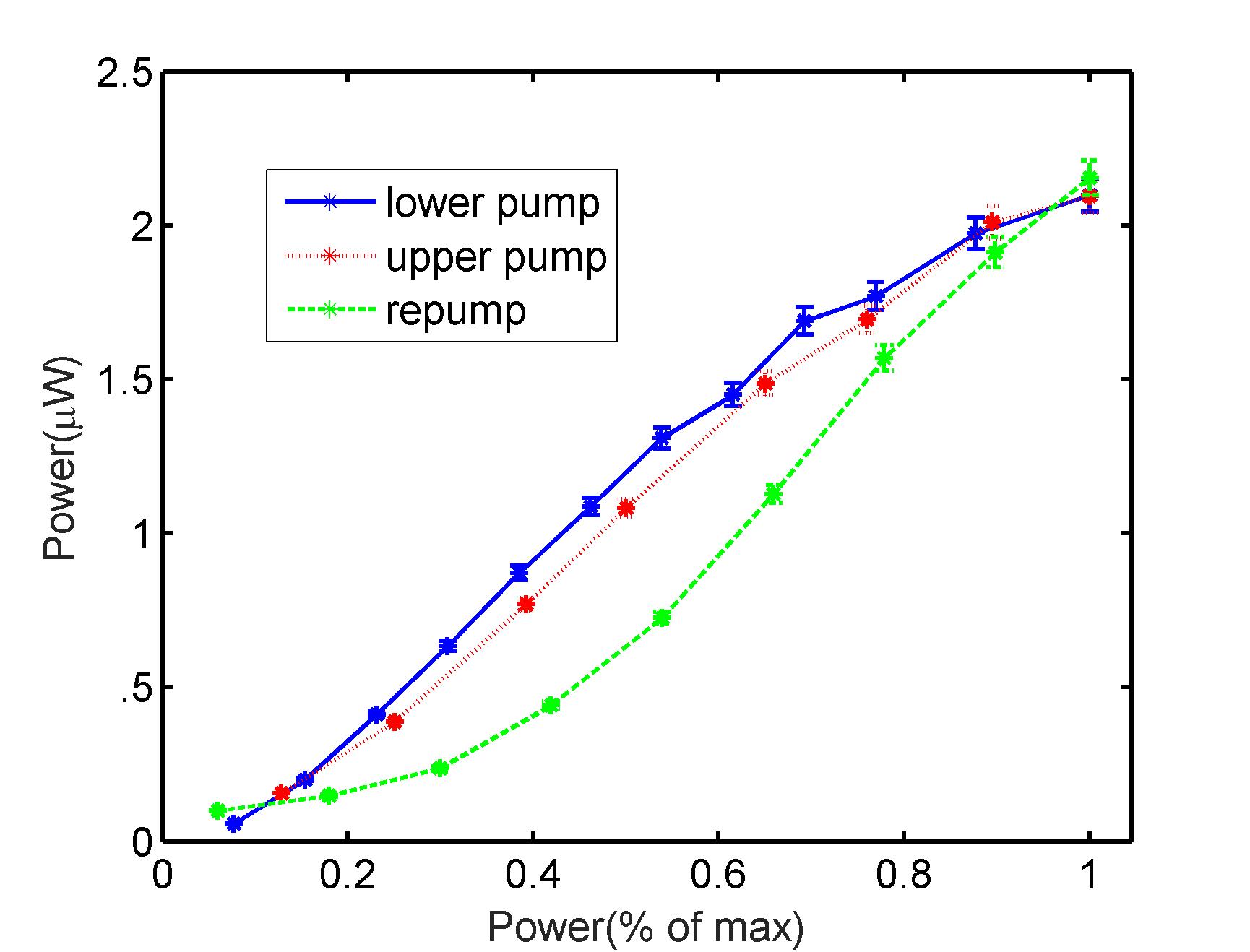}
	\caption{Generated CBL power as a function of normalized power of each pump and repump field. As in Fig.~\ref{fig:D1_blue_on_res}, for each individual measurement the power of one laser was varied between zero and its maximum amount, while the other two lasers were kept at their maximum powers: 65~mW for the lower pump (795~nm), 200~mW for the upper pump (762~nm), and 17~mW for the repump laser (780nm). The laser detuning corresponded to the conditions for the maximum CBL generation in Fig.~\ref{fig:polarization}(c), approximately $+1.2$~GHz blue of the 5$S_{1/2}, F=2\rightarrow 5P_{1/2} $ transition.}
	\label{fig:D1_off_res}
\end{figure}

CBL power dependences were also analyzed for the pump fields detuned  by approximately $+1.2$~GHz from the $5S_{1/2}, F=2\rightarrow 5P_{1/2} $ transition [Fig.~\ref{fig:polarization}(c)]. At this detuning the lower
pump field experienced almost no resonant absorption making the contribution of the step-wise excitation process significantly smaller compared with the direct two-photon excitation. Thus, we observed the maximum blue light generation at the two-photon resonance conditions for the $5S_{1/2} \rightarrow 5D_{3/2}$ transition. We chose to use the linear parallel polarizations arrangement for direct comparison with the resonant case. As one can see in Fig.~\ref{fig:D1_off_res}, in this case the blue light power displays fairly linear dependence on each pump laser field, without reaching saturation or maximum. The repumping power dependence is also qualitatively similar to the resonant case, although it is important to note significantly higher enhancement for the same repump power ($\times 10$ CBL power increase) compare to the resonant case ($\times 4$ CBL power increase).
 
\section{CBL generation via the $D_2$ line}

\begin{figure}[H]
	\centering
	\includegraphics[width=.5\columnwidth]{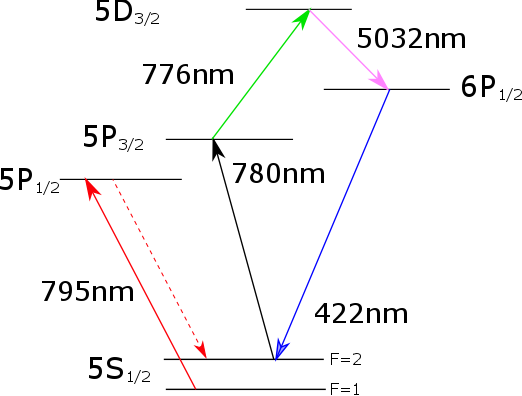}
	\caption{Interaction configuration for CBL generation via the $5P_{3/2}$ intermediate state, that uses the lower pump (780~nm) and the upper pump (776.15~nm) to excite Rb atoms into the $5D_{3/2}$ state, resulting in emission of 5.032~$\mu$m (not experimentally observed) and collimated blue  light (421.7~nm). The repump (795~nm) field is tuned to the hyperfine ground state opposite of the lower pump.}
	\label{fig:levels_D2}
\end{figure}

The alternative excitation pathway to the $5D_{3/2}$ level is through the $5P_{3/2}$ intermediate level. In this case, the two-photon transition was executed using the $D_2$ (780~nm) laser and the Ti:sapph laser, tuned to the 776~nm, while the $D_1$(795~nm) laser served as the repump, as shown in Fig.~\ref{fig:levels_D2}. This pump configuration is traditionally used for the excitation of Rb atoms into the $5D_{5/2}$ state~\cite{zibrov02pra,AkulshinOE09,VernierOE10,bluelightAJP2013,SellOL14,AkulshinOL14}. 
Under the identical experimental conditions, we have obtained up to $120~\mu$W of blue light using the $5S_{1/2}\rightarrow 5P_{3/2}\rightarrow 5D_{5/2}$ excitation scheme, while in the case of the $5S_{1/2}\rightarrow 5P_{3/2}\rightarrow  5D_{3/2}$ pathway the maximum obtained CBL power was just $\le 450$~nW. 
 
\begin{figure}[H]
	\centering
	\includegraphics[width=.7\columnwidth]{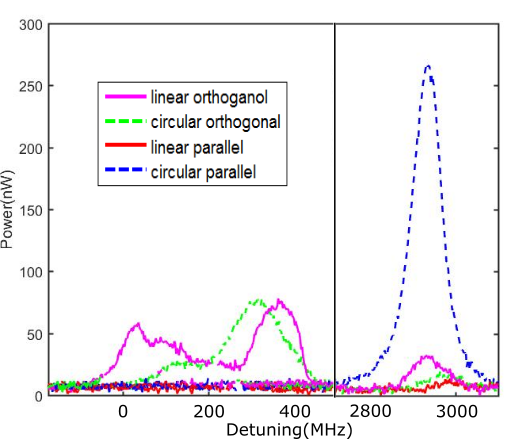}
	\caption{Measured CBL power  for varying polarizatins of lower pump (780~nm) and upper pump (776.1568~nm) as the lower pump is swept across the hyper-fine split  ground states. The considered polarization arrangements for the two pumps are: linear parallel (red, solid line), linear orthogonal (magenta, solid line), circular parallel (blue, dashed line), and circular orthogonal (green, dashed line).}
	\label{fig:D2_polarization}
\end{figure}

Fig.\ref{fig:D2_polarization} demonstrates the measured CBL output power as a function of the lower pump ($D_2$) laser detuning for the previously tested four polarization combinations, shown in Fig.~\ref{fig:polarization}. We observed an even more pronounced dependence of the blue light power on the pump polarizations than in the $D_1$ excitation scheme. For the two-photon $5S_{1/2}, F=2\rightarrow 5D_{3/2}$ transition, the orthogonally circularly-polarized pump fields yielded CBL output that was stronger than the other three configurations by at least an order of magnitude. Remarkably, the same pump polarization arrangement produced no CBL when the lower pump was tuned to the other hyperfine ground state $5S_{1/2}F=3$. At that frequency the blue light was detected only for the circular orthogonal and linear orthogonal polarizations. Overall, we found significantly weaker (approximately by a factor of 10) blue light generation, compare to the $D_1$ excitation scheme. Also,  the blue light power dropped very rapidly with the laser detuning away from the resonance, so that no detectable CBL output was found at $+1.2$~GHz detuning used for the off-resonant case in previous section. 

\begin{figure}[H]
	\centering
	\includegraphics[width=.7\columnwidth]{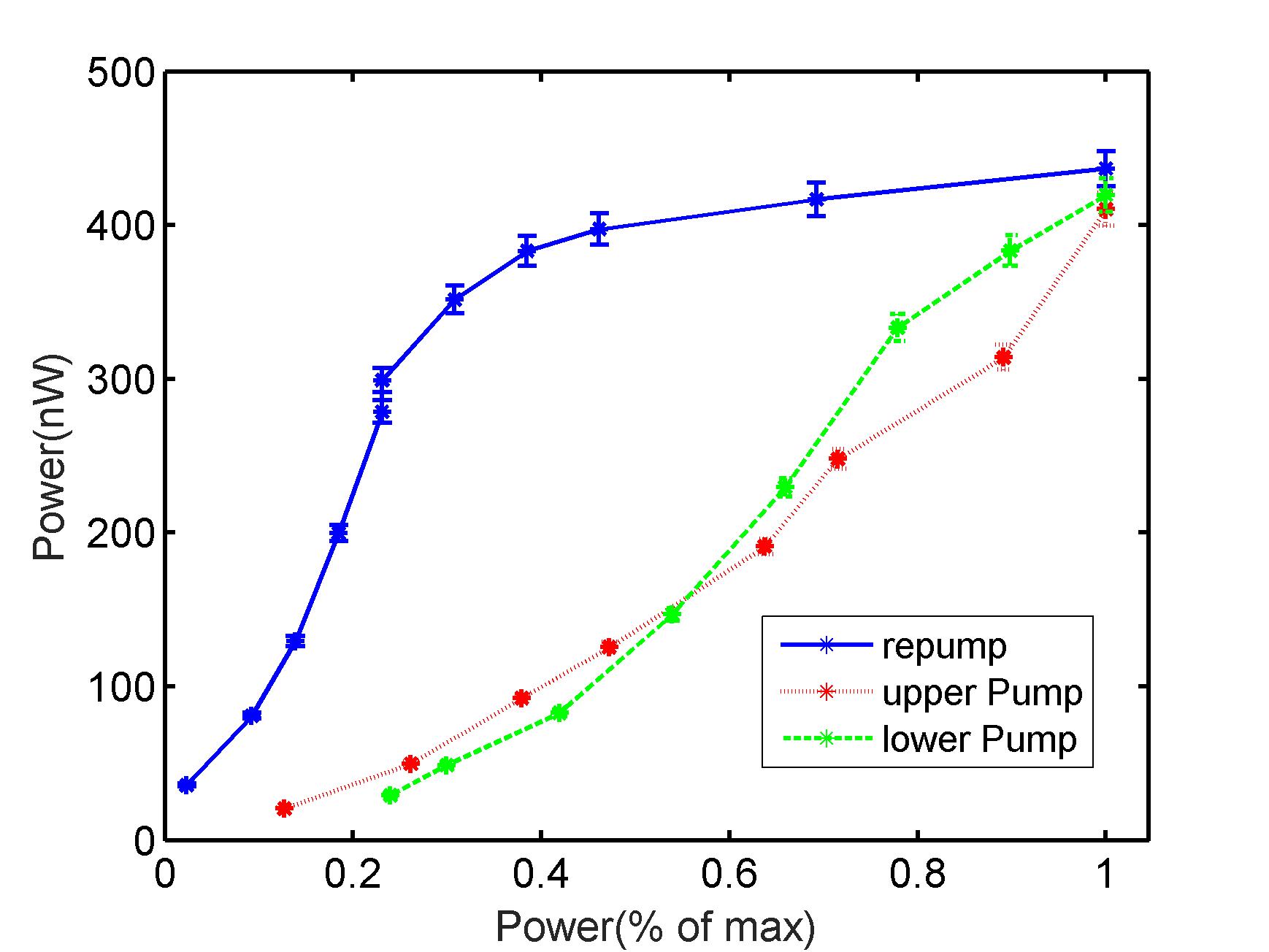}
	\caption{Generated CBL as a function of normalized power of the pump and repump fields. As in Fig.~\ref{fig:D1_blue_on_res}, for each individual dependence, the power of one laser was varied between zero and its maximum value, while the other two lasers were kept at their maximum powers: 17~mW for the lower pump (780~nm), 200~mW for the upper pump (776~nm), and 65~mW for the remupmer (795~nm). The laser detuning corresponded to the conditions for the maximum CBL generation in Fig.~\ref{fig:D2_polarization}(b), near $S_{1/2}F=2\rightarrow 5P_{3/2}F^\prime$ transition. The upper pump wavelength was fixed at $776.1568$~nm.}
	\label{fig:D2_blue}
\end{figure}

Fig.~\ref{fig:D2_blue} shows the dependences of the CBL power on the power of the pump and the repump lasers, measured for parallel circular polarization of the pumps, the configuration yielding the highest CBL powers. When either pump power was varied, we observed a roughly linear dependence for the blue light output. Unlike the resonant excitation using the $D_1$ optical transition shown in Fig.~\ref{fig:D1_blue_on_res}, no signs of saturation or peaking was observed at the available range of the lower pump ($D_2$ laser) power. It is important to note, however, that we operated with less available laser power. In the case of the $D_1$ excitation channel, the CBL power started to saturate at around $12$~mW of the lower pump power, reaching the maximum value at $35$~mW. Since the maximum available $D_2$ laser power was only $17$~mW, it is possible that nonlinear power dependence can be observed at higher pump powers.  

The repump power dependence shows clear saturation for the $D_1$ laser powers above $\approx 20$~mW, the power level necessary to provide efficient depopulation of the $5S_{1/2} F=3$ ground state. Unsurprisingly, further repump power increase did not provide any additional advantages. We confirmed this by the additional measurements of the $D_2$ laser resonant absorption, observing an increase in absorption from $30\%$ without the repump to a plateau of $\approx 50\%$ with repump power above $20$~mW. 
%

\section{Simplified theoretical simulations} \label{sec:theory}

To gain some qualitative understanding of the observed experimental behavior, we built a simplified theoretical model of the blue light generation using the methodology described in Ref.~\cite{mikhailov2013jmo_fast_N_scheme} adopted for in a four-level diamond scheme. To reduce the complexity, we have neglected the nuclear spin, eliminating the hyperfine structure. To account for alternative spontaneous decay paths and the optical pumping of atoms in the second ground hyperfine state, we introduced an additional fictional non-degenerate ground state. Lifetime and branching ratios of which, match those of the corresponding Rb states.
We also do not account for the Doppler broadening of the optical transition due to the thermal motion of the atom, but incorporate the ground-state decoherence rate of $1$~MHz, mimicking the transient relaxation. 

\begin{figure}[H]
	\centering
	\includegraphics[width=1.0\columnwidth]{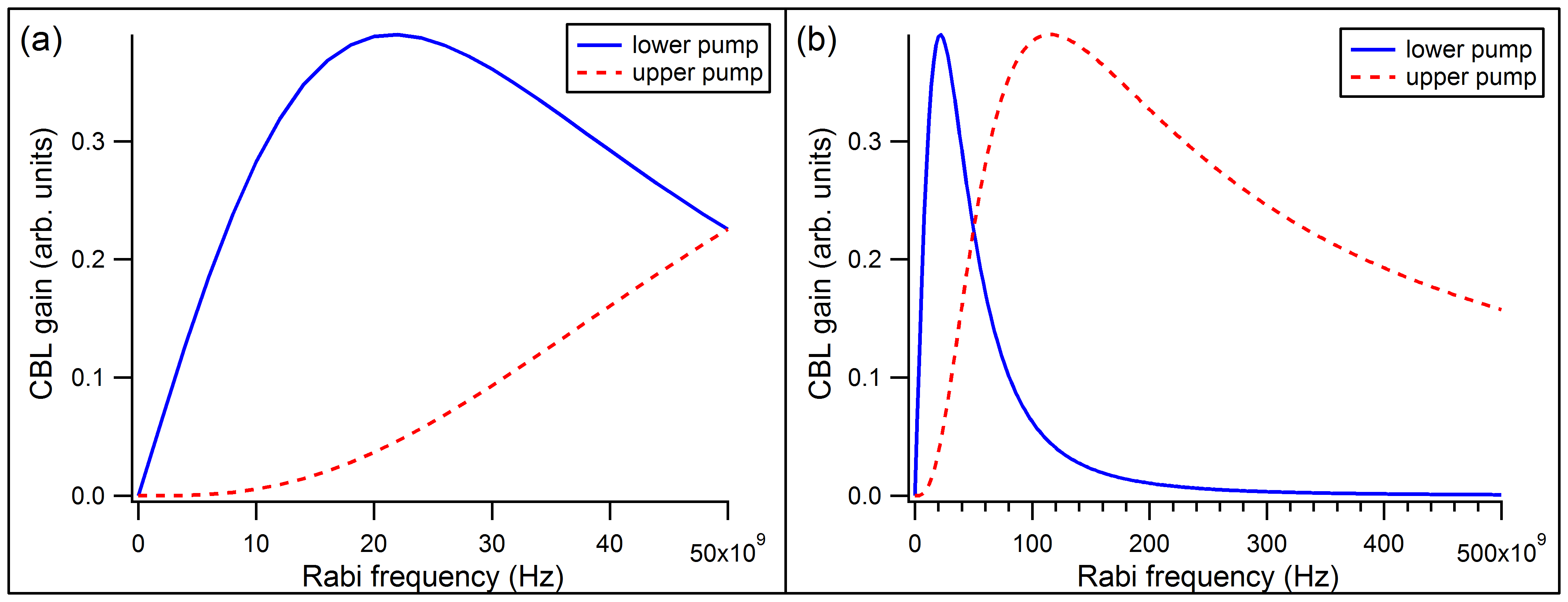}
	\caption{Calculated CBL gain as a function of either pump Rabi frequency. While the Rabi frequency of one of the pump fields is varied, the other is maintained at its maximum value of $5\times 10^{10}$~Hz. In \emph{(a)} the Rabi frequencies change in the range similar to those used for experimental data in Fig.~\ref{fig:D1_blue_on_res}. In \emph{(b)} the range of variation is increased by a factor of 10 to display the more complete power dependence. For these simulations we used parallel circular polarizations for all optical fields; however, the same general behavior is observed for other polarization configurations. }
	\label{fig:power_theory}
\end{figure}

Despite many simplifications, the calculations qualitatively match the experimental observations and provide explination for the observed behaviors. Fig.~\ref{fig:power_theory}(a) demonstrates the dependences of the CBL gain on the powers of the pump lasers in the range of Rabi frequencies comparable with those used in the experiment. The simulated trends are similar to the experimental dependencies, shown in Fig.~\ref{fig:D1_blue_on_res}, in which the CBL power grows with the upper pump, but reaches a maximum and then declines when the lower pump power is increased. However, if we allow either pump power to vary at a larger range, as shown in Fig.~\ref{fig:power_theory}(b), we see that the maximum CBL output occurs when the upper to lower pump Rabi frequencies ratio is roughly $2.3$. Increase of either pump power leads to reduction of populations of the atomic levels, involved in blue light generation and consequently to the reduction of CBL output.  

This understanding also help to explain the difference in the CBL output dependence on the lower pump power at $D_1$ and $D_2$ lines, shown in Figs.~\ref{fig:D1_blue_on_res} and \ref{fig:D2_blue}. Since the D1 laser output is higher, we were able to realize the optimal power ratio for the lower and upper pumps and observed the CBL maximization. However, if the maximum value of the D1 pump was used, we were not able to reach the optimal CBL conditions due to power limitation of the Ti:Sapph laser. However, when we tested the alternative configuration through the $5P_{3/2}$ state, the mismatch between the available powers of the two pumps restricted the CBL generation to the lower part of the theoretical curve.          

\begin{figure}[H]
	\centering
	\includegraphics[width=0.6\columnwidth]{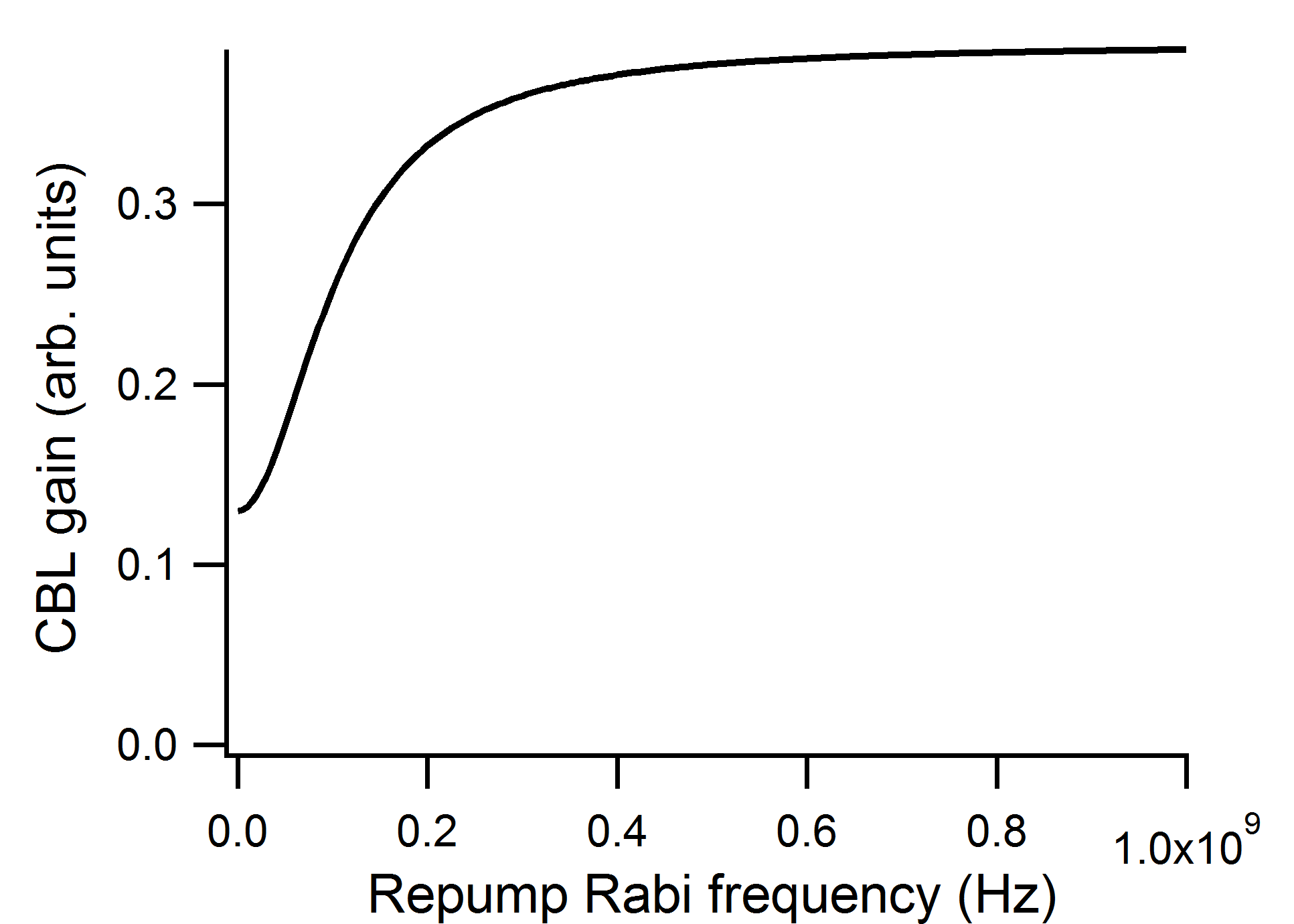}
	\caption{Calculated CBL gain as a function of repump Rabi frequency. For these simulations we used parallel circular polarizations for all optical fields, and the Rabi frequencies of the lower and upper pump fields of $2\times 10^{10}$~Hz and $5\times 10^{10}$~Hz, corresponding to the calculated maximum CBL gain. }
	\label{fig:repump_theory}
\end{figure}

We also calculated the dependence of CBL yield on the repump laser strength. As expected, efficient repumping of atomic population from uncoupled ground state magnifies the CBL gain significantly, reaching saturation. This is qualitatively the same behavior as observed experimentally in Fig.~\ref{fig:D2_blue}, when a more powerful D1 laser served as a repumper. Because of the lower maximum available output of the D2 laser and its stronger resonant absorption, we did not achieve such saturation when it was used for repumping, and the corresponding line at Fig.~\ref{fig:D1_blue_on_res} resembles the lower end of the simulated curve.

\begin{figure}[H]
	\centering
	\includegraphics[width=1.0\columnwidth]{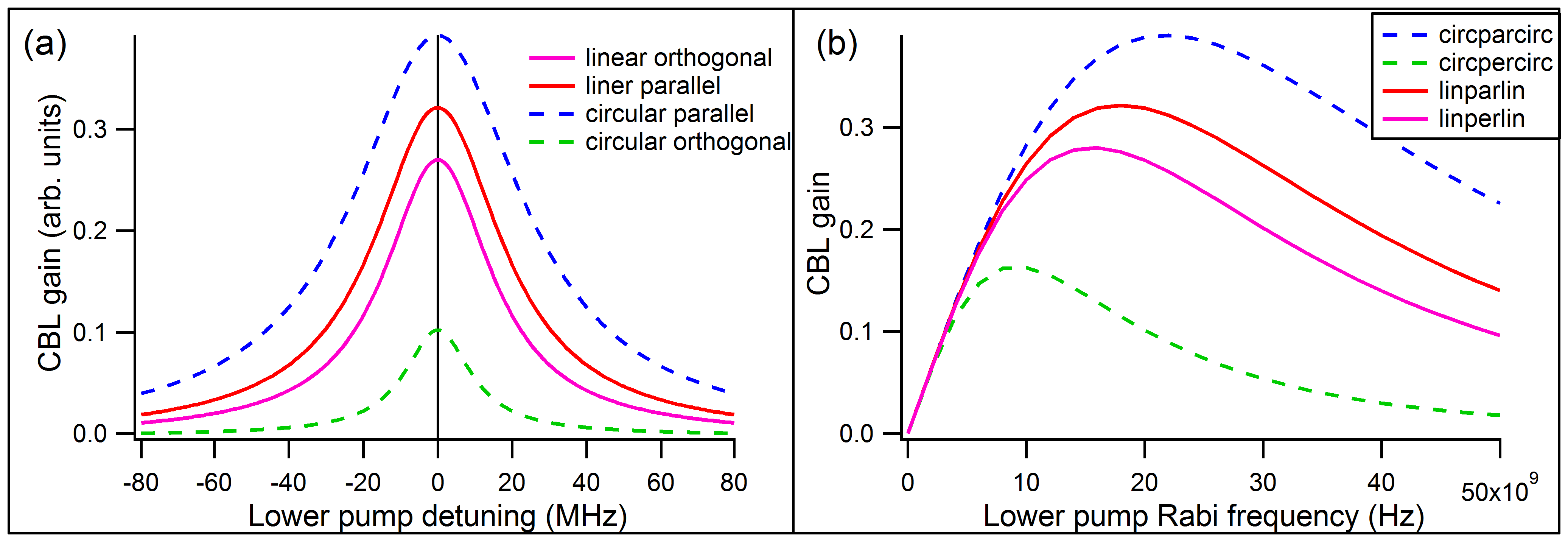}
	\caption{\emph{(a)} Calculated CBL gain as a function of lower pump frequency for the four polarization arrangements tested in the experiment. For these simulations the Rabi frequencies of the lower and upper pump fields of $2\times 10^{10}$~Hz and $5\times 10^{10}$~Hz, and the upper pump was resonant with the corresponding optical transition. \emph{(b)} Modification of the CBL gain lower power dependence for different polarization arrangements. The simulation parameters are identical to those using in Fig.~\ref{fig:power_theory}.}
	\label{fig:polar_theory}
\end{figure}

Finally, we can check the effect of the polarizations of the pump fields. Fig.~\ref{fig:polar_theory} presents the results of the simulations of CBL gain for the four polarization configurations tested in the experiment. While inclusion of accurate Zeeman and hyperfine atomic structure is necessary to match the experimentally measured dependences, the simplified simulations still display some general features, characteristic to the observations. For example, in the simulations for both linearly and circularly polarized pump fields, larger CBL gain is observed when the two pumps have parallel, rather than orthogonal. We also verified that the observed changes in CBL strength for different polarizations is general, and not specific for particular values of pump powers. For that we replicated the CBL gain dependence on the lower pump Rabi frequency, shown in Fig.~\ref{fig:polar_theory}(b).

\section{Conclusion}

In conclusion, we report on characterization of the collimated blue light generation via two-photon excitation from the ${}^{85}$Rb $5S_{1/2}$ ground state to the $5D_{3/2}$ excited state through either $5P_{1/2}$ or $5P_{3/2}$ intermediate levels. We have studied the characteristics of the generated blue light for various pump laser frequencies, and found that the polarization arrangement leading to the maximum CBL power output strongly depends on the optical transitions used. This indicates the importance of selection rules and individual Zeeman transition probabilities. The experimental results shared various qualitative characteristics with the theoretical simulation. We  found that under the optimized experimental conditions the blue light output was noticeably stronger when the $D_1$ optical transition was used as the first excitation step. In the case of the $D_1$ resonant excitation we demonstrated the existence of the optimal pump powers that led to maximum blue output. For other situations (off-resonant $D_1$ excitation or resonant $D_2$ excitation) a linear dependence of output CBL power on the lower pump power was detected. Theoretical simulations allow us to explain this behavior: for each set of experimental parameters there seems to be an optimal ratio between the lower and upper pumps that lead to maximum CBL yield. Any deviations from this value result in sub-optimal population redistribution between the involved atomic transitions, and in the reduction of blue light generation. In case of the $D_1$ resonant excitation we were able to realize such optimal conditions for the lower pump power. For the other configurations, however, we were limited to the initial rising power dependence, before the CBL maximum was reached. Our measurements and simulation also demonstrated the importance of the repumping of atomic population from the uncoupled ground state sublevels, that led to an order of magnitude increase in blue light generation in all tested configurations.  A more detail simulation and further study may shed light on the specific temperatures, powers, and polarizations for a better optimized and efficient CBL generation.

\section{Aknowledgenets}

This research was supported by the National Science Foundation grant PHY-308281. We would like to thank S. Rochester for making the Mathematica package AtomicDensityMatrix available on-line that enabled us to carry out the numerical simulations.


\end{document}